# Calibration of the Spectrometer aboard the INTEGRAL satellite

Stéphane Schanne[1a], Bertand Cordier[a], Maurice Gros[a], David Attié[a],
Peter v. Ballmoos[b], Laurent Bouchet[b], Raffaelo Carli[e], Paul Connell[h], Roland Diehl[c], Pierre Jean[b],
Jürgen Kiener[d], Andreas v. Kienlin[c], Jürgen Knödlseder[b], Philippe Laurent[a], Giselher Lichti[c],
Pierre Mandrou[b], Jacques Paul[a], Philippe Paul[b], Jean-Pierre Roques[b], Filomeno Sanchez[i],
Volker Schönfelder[c], Chris Shrader[g,f], Gerry Skinner[b], Andy Strong[c], Steve Sturner[g,f],
Vincent Tatischeff[d], Bonnard Teegarden[f], Gilbert Vedrenne[b],
Georg Weidenspointner[g,f], Cornelia Wunderer[c].

[a] DAPNIA/Service d'Astrophysique, CEA/Saclay, Bâtiment 709, 91191 Gif-sur-Yvette, France
[b] Centre d'Etudes Spatiales des Rayonnements, 9, av. du Colonel Roche, 31028 Toulouse, France
[c] Max-Planck-Institut für extraterrestrische Physik, Giessenbachstraße, 85748 Garching, Germany
[d] Centre de Spectrométrie Nucléaire et de Spectrométrie de Masse, IN2P3-CNRS, 91405 Orsay Campus, France
[e] European Space Agency/ESTEC, Keplerlaan 1, 2200 AG Noordwijk, The Netherlands
[f] NASA/Goddard Space Flight Center (Code 661), Greenbelt, MD 20771, USA
[g] Univ. Space Research Association, 7501 Forbes Blvd., Seabrook, MD 20706, USA
[h] Space Research Group, Univ. of Birmingham, Birmingham B15 2TT, UK
[i] IFIC, Univ. of Valencia, 50 av. Doctor Moliner, 46100 Burjassot, Spain

## ABSTRACT

SPI, the Spectrometer on board the ESA INTEGRAL satellite, to be launched in October 2002, will study the gamma-ray sky in the 20 keV to 8 MeV energy band with a spectral resolution of 2 keV for photons of 1 MeV, thanks to its 19 germanium detectors spanning an active area of 500 cm$^2$. A coded mask imaging technique provides a 2° angular resolution. The 16° field of view is defined by an active BGO veto shield, furthermore used for background rejection. In April 2001 the flight model of SPI underwent a one-month calibration campaign at CEA in Bruyères le Châtel using low intensity radioactive sources and the CEA accelerator for homogeneity measurements and high intensity radioactive sources for imaging performance measurements. After integration of all scientific payloads (the spectrometer SPI, the imager IBIS and the monitors JEM-X and OMC) on the INTEGRAL satellite, a cross-calibration campaign has been performed at the ESA center in Noordwijk. A set of sources has been placed in the field of view of the different instruments in order to compare their performances and determine their mutual influence. Some of those sources had already been used in Bruyères during the SPI standalone test. For the lowest energy band calibration an X-ray generator has been used. We report on the scientific goals of this calibration activity, and present the measurements performed as well as some preliminary results.

**Keywords:** Space Telescope, Germanium detectors, Gamma-ray Astronomy, Calibration, INTEGRAL, SPI.

---

[1] Further author information: Schanne@hep.saclay.cea.fr; phone +33 1 69 08 15 47; fax + 33 1 69 08 79 96. Comissariat à l'Energie Atomique, CEA Saclay; Département d'Astrophysique, de Physique des Particules, de Physique Nucléaire et de l'Instrumentation Associée; 91191 Gif-sur-Yvette, France. http://www-dapnia.cea.fr

# 1. OVERVIEW

**1.1. The INTEGRAL satellite**

In 1993 the European Space Agency (ESA) selected the International Gamma Ray Astrophysics Laboratory INTEGRAL [1] as the next gamma-ray astronomy mission in the 15 keV to 10 MeV energy domain, successor to NASA's Compton Gamma Ray Observatory and the Russian/French GRANAT/SIGMA missions. The scientific goals of INTEGRAL cover a wide range of astrophysics topics [2], including the study of:

- Nearby supernovae, if such star explosions should occur during the mission's lifetime.
- Stellar nucleosynthesis, via detection and cartography of short-, medium- and long-lived nuclides like $^{56}$Co, $^{44}$Ti, $^{26}$Al and $^{60}$Fe.
- Structure of our galaxy, including a regular galactic plane scan for the detection of transient sources and for detailed studies of the galactic center.
- Compact Objects, identification of high-energy sources, particle acceleration processes and extra-galactic gamma ray astronomy.

The key parameters of INTEGRAL in order to reach these goals are:

- Energy resolution of about 2‰ and narrow-line sensitivity of about $5\times10^{-6}$ photons s$^{-1}$ cm$^{-2}$ at 1 MeV for a 3σ detection after an exposure time of $10^6$ s, achieved with the INTEGRAL Spectrometer (SPI) [3-6].
- Angular resolution of 12' and source localization of about 1', achieved with the INTEGRAL Imager (IBIS) [7].
- Two additional instruments permit the monitoring of the INTEGRAL field of view down to lower photon energies, the X-ray monitor (JEM-X) and the optical camera (OMC).

INTEGRAL will be launched on October 17, 2002, using the Proton launch vehicle of the Russian Space Agency from the launch site in Baïkonur (Kazakhstan). It will have an eccentric orbit with an apogee of 153000 km and a 72 h orbiting period. During its two (up to five) years mission it will be operated as an international observatory with 70% of the observing time allocated to external observers and 30% to the instrument collaborations.

**1.2. The spectrometer onboard INTEGRAL**

The spectrometer SPI is built by an international collaboration of research institutes under technical leadership of the CNES, the French Space Agency. It is 238 cm high and has a mass of 1300 kg (see Fig. 1). SPI features a position sensitive detector (γ-camera) mounted below a hexagonal coded mask, surrounded by an active veto shield.

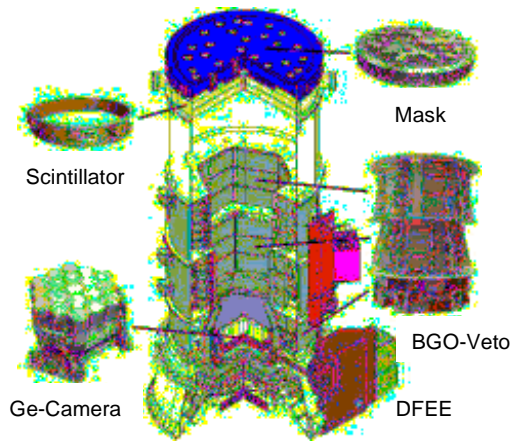

**Figure 1.** Elements of the Spectrometer aboard the INTEGRAL satellite.

The gamma camera is composed of a hexagonal array of 19 high purity germanium detectors covering an area of 500 cm$^2$ kept at a temperature of 85 K by an active cryogenic system. It provides an energy resolution of about 2 keV for photons of 1 MeV, thus has a 10 times better narrow-line sensitivity than previous telescopes. During flight, due to cosmic ray interactions in the germanium crystals, the resolution (thus the sensitivity) can degrade with time. In this case, in order to repair the crystal structure, it is foreseen to heat the crystals up to 100° C for 24 hours and thus "anneal" the crystal defects.

A mask of tungsten blocks is placed 171 cm above the Ge-camera such that the mask throws a shadow onto the detection plane. Image deconvolution tools permit the reconstruction of the sky image with ~2° resolution. The fully coded field of view of SPI covers 16°. In order to achieve a good imaging performance, the spectrometer takes subsequent images of the same field of view under a slightly different pointing angle. Every 30 mn, using inertial wheels, the spacecraft is therefore reoriented following a "dithering" pattern.

For efficient background rejection the detector is surrounded by a massive anti-coincidence system (ACS) made of 91 BGO scintillation detector crystals viewed by photo-multipliers. A 5 mm thick plastic scintillator (PSAC) placed below the mask permits the rejection of events from prompt mask activation.

A Pulse Shape Discriminator (PSD) classifies the γ-photons impinging on the germanium detectors. Above 300 keV the photon interaction is dominated by Compton scattering, identified as multi-pulse events, for which inside a detector crystal more than one photo-electron is produced per impinging γ. This allows rejection of single-pulse events, mainly due to background radioactive decays in the detector crystal itself.

The Digital Front End Electronics (DFEE) provides event timing with 100 μs accuracy, event building and classification, event rejection using the anti-coincidence veto signal, and event counting and dead-time monitoring in order to determine absolute incident photon intensities.

**1.3. The SPI data acquisition system**

Each of the 19 Ge-detector crystals is subject to an electric potential of 4000 V between the outer side of the crystal and the central anode. An impinging photon extracts from the valence band of the semiconductor one or more electrons, which migrate in the electric field, inducing an electrical current in the anode. This analogue signal is pre-amplified and sent both to the PSD for pulse shape analysis and to the Analogue Front End Electronics (AFEE), which emits a Time Tag (TT) to the DFEE synchronously with the event (see Fig. 2). In the AFEE the analogue signal is then integrated and converted by an ADC system to a 15-bit numerical value (1 bit for low or high energy gain classification and 14 ADC bits) and sent asynchronously after about 27 μs to the DFEE via a serial transmission protocol.

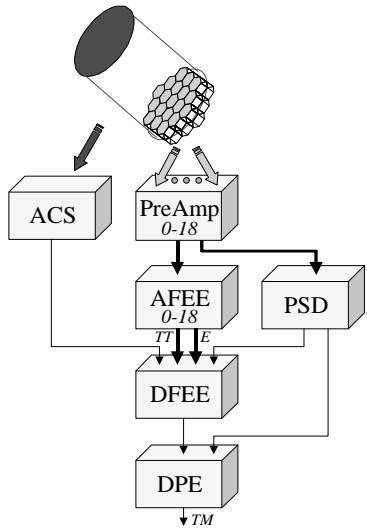
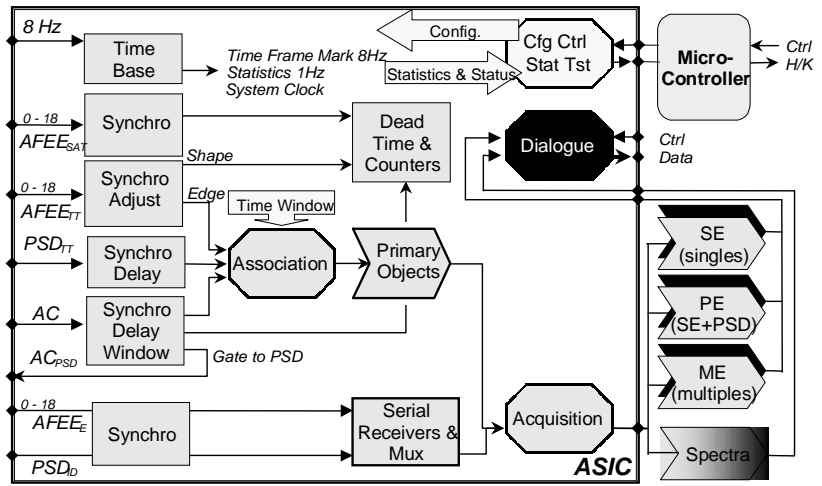

**Figure 2.** The SPI data acquisition.   **Figure 3.** Block diagram of the SPI-DFEE with its digital ASIC.

The front stage of the DFEE (see Fig. 3) receives the ACS veto signal (dead-time of 725 ns per hit), the 19 time tags sent by the AFEE and a time tag sent by the PSD for every analyzed event. Each signal is delayed internally by a time value adjustable in steps of 50 ns.

The output signals of the DFEE front stage are used by the association machine of the DFEE in order to perform the event building. A time tag is considered as a Single Detector Event (SE) if no other time tag occurs within the duration of an adjustable "association time window" (typically 350 ns). The event is a Multiple Detector Event (ME) if a subsequent time tag arrives before the closure of the "association time window", opened by a previous time tag. A SE with one PSD-Time Tag arriving during the "association time window" is considered as a Single Detector Event with PSD information (PE). The absolute time of an event classified as SE, ME or PE is measured by the DFEE in units of 102.4 µs with respect to an onboard 8 Hz clock. Within a ME the time between subsequent time tags is measured in units of 50 ns. All events are stored into an internal FIFO in order to wait for the arrival of the numerical values of the energies determined by the AFEE, after which they are extracted from the FIFO and written to output tables together with their energy values. Those tables are handled in a dead-time free double-buffer mode and sent on the basis of a time frame defined by the onboard 8 Hz clock to the Digital Processing Unit (DPE). The DPE assembles the PE information with the result of the PSD analysis and sends the resulting stream to the satellite telemetry downlink system.

The high speed data processing of the DFEE is entirely handled by a digital ASIC [8-9] designed at CEA Saclay and built by TEMIC/Atmel in a radiation tolerant CMOS technology required by the space environment (maximal integrated dose 0.3 Gy).

## 2. INTEGRATION AND TEST CAMPAIGNS OF SPI

The SPI flight model was fully integrated and tested between September 2000 and March 2001 at CNES in Toulouse (France). The scientific performance tests showed that the instrument meets its design goals. The precise timing alignment within 50 ns of the 19 detector channels with respect to the Pulse Shape Discriminator and Anti-Coincidence signals was successfully performed using the built-in test mechanisms of the Digital Front End Electronics.

In March 2001, placed inside the vacuum chamber at CNES, SPI underwent its thermal vacuum test campaign, where the instrument was operated under close-to-space conditions. The Ge detectors were cooled down to 85 K using the four onboard cryo-cooling machines. Different cryo-cooler operation modes were exercised in order to study their behavior. Measurements using a set of radioactive sources showed that the spectral resolution obtained with cold preamplifiers is better than the design goal of 2‰. For example using a $^{60}$Co source, the measured 1332 keV peak showed a full width at half maximum (FWHM) between 2.1 and 2.6 keV for all 19 detectors over a large temperature range, confirming that a good energy resolution is expected in flight in case of low radiation damages or after annealing.

During the month of April 2001 the SPI on-ground calibration campaign [10-11] took place at CEA in Bruyères le Châtel (France). In May 2001 the spectrometer was delivered to ESA and was integrated on board of the satellite at Alenia in Torino (Italy).

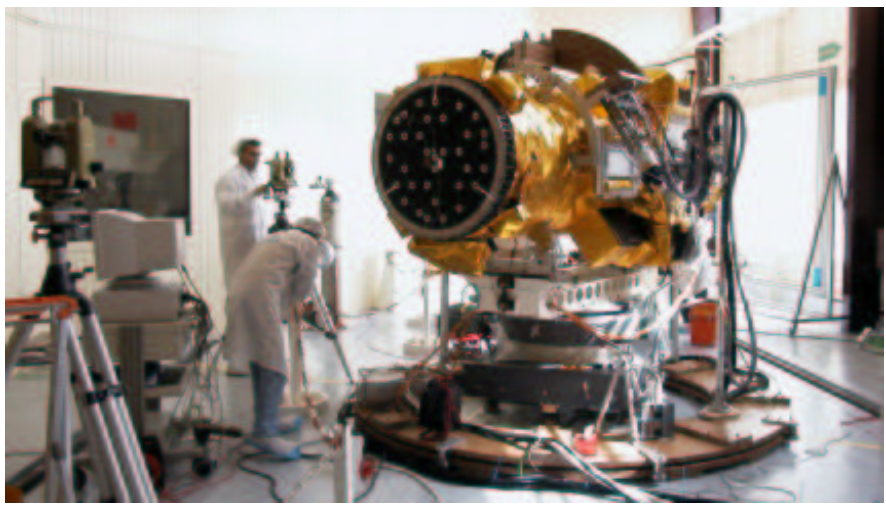

**Figure 4.** The INTEGRAL spectrometer Flight Model on its rotating on-ground support at the Bruyères le Châtel calibration site during installation. The mask structure is visible, the mask blocks are located inwards. The accelerator target is located behind the left window. The long distance sources are located outside the clean room at 125 m from SPI, behind the XY-beam profile scanner visible in front of the right window.

In February 2002, after completion of the INTEGRAL assembly at ESA in Noordwijk (The Netherlands), a cross-calibration between all INTEGRAL instruments took place. A series of end-to-end tests were performed in January and

June 2002, with data transmission from INTEGRAL to the MOC (Mission Operation Center) at ESA/ESOC in Darmstadt (Germany) and the ISDC (INTEGRAL Science Data Center) in Geneva (Switzerland). In May 2002, the INTEGRAL satellite was placed in the large vacuum chamber at ESA in Noordwijk, where thermal cycles during orbit were simulated. Using radioactive sources placed inside the chamber, the SPI detector resolution and timing alignment was checked again and the PSD calibration was performed. The satellite will be transferred to Baïkonur during summer 2002, where functional checks will be performed prior to launch.

## 3. ON-GROUND CALIBRATION OF SPI

The SPI on-ground calibration campaign [10] was performed during the whole month of April 2001 at the CEA center of Bruyères le Châtel (France). This site was selected since it offered the possibility for the manipulation of high activity radioactive sources as well as the use and modification of its accelerator complex. A clean room experimental hall with an airlock annex was built in order to accommodate the space equipment close to the accelerator. The calibration lasted 21 days (including 108 h of accelerator measurements) and was operated continuously, with a turnaround of three shifts per day by 38 scientists from seven laboratories.

The aim of this measurement campaign was to determine the detector efficiencies, the homogeneity of the detection plane, and the pre-launch imaging performance for the whole energy range of 20 keV to 8 MeV. Therefore a series of measurements was taken with sources at different angles of incidence and at distances of 8 m for the energy and efficiency calibration, and 125 m for the imaging calibration.

SPI was installed with its optical axis in the horizontal position on a motorized mechanical on-ground support permitting a precise 360° rotation around its vertical axis and two positions around the optical axis of SPI, such that the sources could be seen by SPI at different angles (Fig. 4). The whole field of view could be monitored and different spacecraft dithering points simulated. During the SPI on-ground calibration a total of 413 data acquisition periods (runs) in different configurations were recorded and are currently being analyzed.

### 3.1. Efficiency and homogeneity measurements in the low-energy domain

The efficiency and homogeneity measurements in the energy domain below 2 MeV were performed using 11 different radioactive sources (each with activities between 5 and 12 MBq) producing photons in the range from 60 keV ($^{241}$Am source) to 1836 keV ($^{88}$Y source). In subsequent acquisition periods those sources were placed on the optical axis of SPI at a distance of 8 m. In order to illuminate all detectors uniformly the spectrometer's mask was removed.

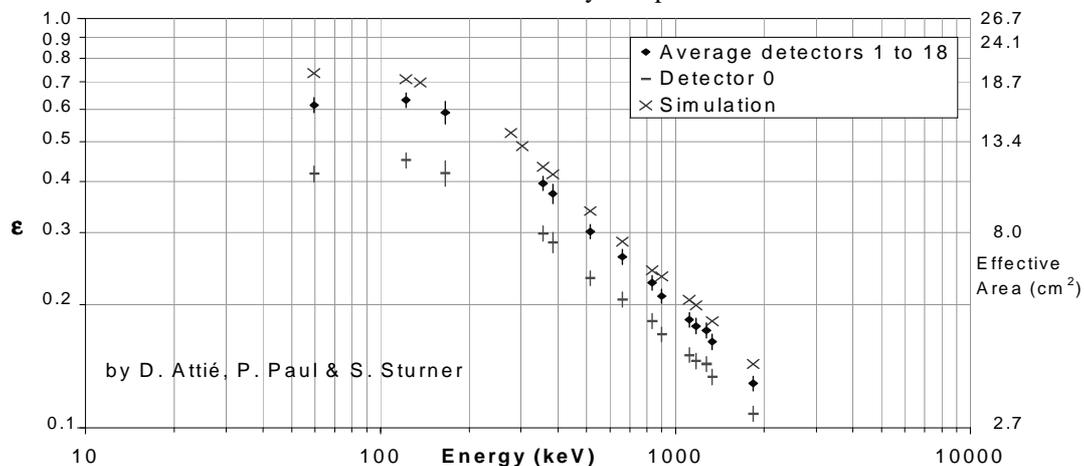

**Figure 5.** Overall photopeak efficiency $\varepsilon$ as a function of energy for single detector hits (SE+PE), obtained for an alignment $\alpha=0°$ without mask. In this configuration detector 0 is shadowed by the central PSAC alignment device, hence its lower efficiency. 3$\sigma$ error bars are plotted. The GEANT Monte-Carlo simulation is over-predicting the measured efficiency by ~11%, although the simulations reproduce nicely the peak and the Compton tail line shapes.

Preliminary results are presented in Fig. 5, which shows the overall SPI detector photoelectric-peak-efficiency as a function of energy (mask dismounted). In order to derive this efficiency, for each γ-ray line of a radioactive source used, the photon flux impinging on a detector is computed using the known source activity, the branching ratio of the given line, the air transmission coefficient at this energy and the distance between the source and the detection plane (8.2 m).

From the computed flux, the expected number of photons hitting each detector is evaluated, taking into account the effective detection area for each of the 19 Ge detectors (mean effective area ~ 26.7 cm$^2$) and the duration of each measurement, corrected by the instrumental dead-time (mean value ~ 2%, dominated by the ACS dead-time, measured online by the DFEE for each detector). In order to derive the efficiency, this expected number of photons is compared to the number of photons resulting from the data by fitting each photoelectric peak with the sum of a Gauss-function, a step-function and a straight line, after background subtraction. The absorption is not corrected for material in front of the Ge detectors except the mask (like the PSAC and the beryllium cryostat), and contributes to the presented overall detector efficiency.

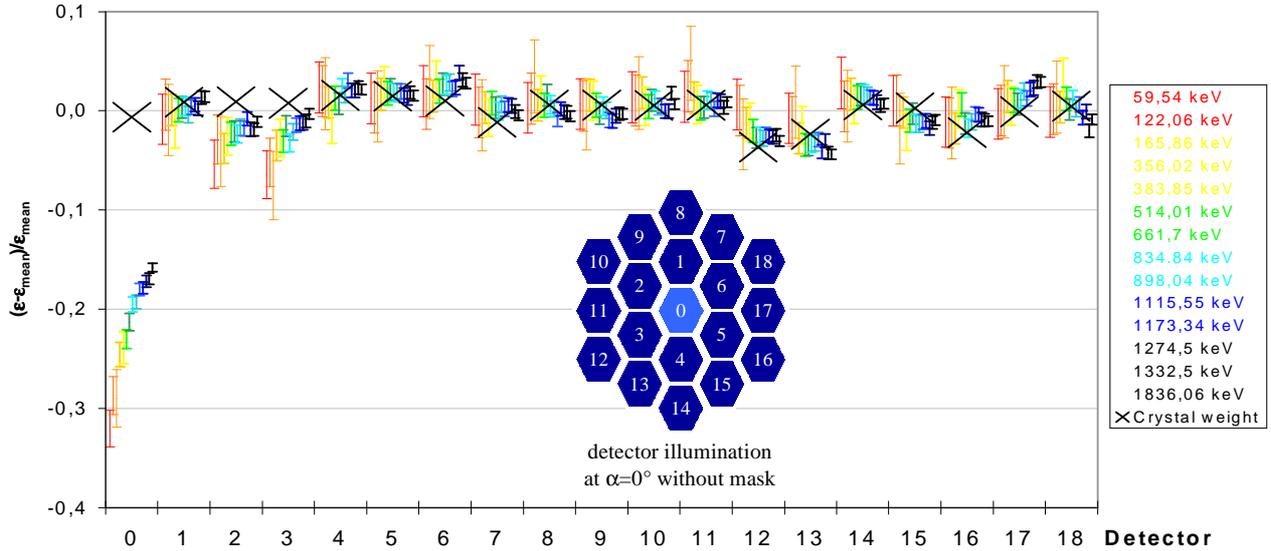

**Figure 6.** The detection plane homogeneity (measured at $\alpha=0°$) is defined as the individual detector efficiency compared to the mean efficiency of detectors 1 to 18. Detector 0 is shadowed by the PSAC-alignment device. The detector numbering is defined in the inset. For each detector the energy dependence of the homogeneity is plotted (for a given detector, the line energy increases from left to right according to the displayed list of energy lines from short distance radioactive sources). The crosses represent the homogeneity of the germanium crystal weights.

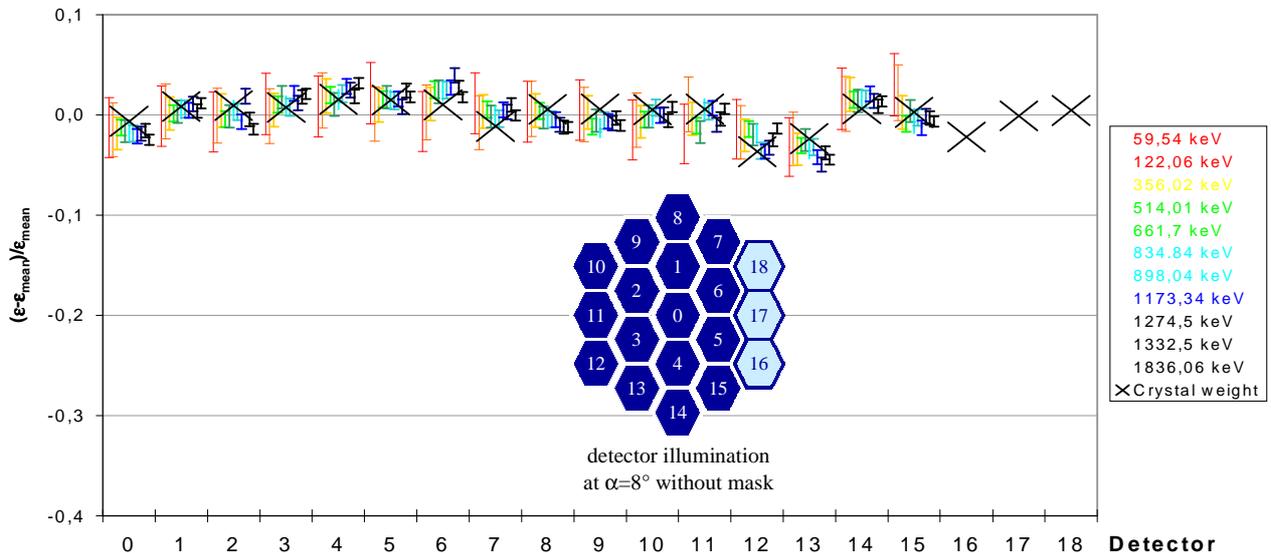

**Figure 7.** The detection plane homogeneity (measured at $\alpha=8°$) defined as the individual detector efficiency compared to the mean efficiency of detectors 0 to 15. Detectors 16 to 18 are shadowed by the ACS. The alignment device is projected outside the detection plane. The efficiency non-homogeneity follows that of the crystal weights.

Fig. 6 shows the homogeneity of the detection plane, together with its energy dependence on the same figure, for sources aligned with the optical axis of the spectrometer (angle between the optical axis and the source α=0°). The homogeneity is computed as the deviation from the mean efficiencies of the detectors 1 to 18, central detector 0 excluded. In this configuration, the lower efficiency measured at detector 0 is due to the presence of an alignment device between the detector and the source. This device has been incorporated at the center of the PSAC, where a mechanically stabilized hole has been drilled, with a corresponding viewfinder at the top of the beryllium cryostat. At the angle α=0°, this device is aligned with the detector 0, reducing its overall efficiency by as much as 30% at 60 keV.

With the same sources, data have been taken at an alignment angle α=8° between sources and optical axis. In this configuration the alignment device in the PSAC is projected outside the detection plane and the overall efficiency of detector 0 is recovered, while detectors 16 to 18 show a very low efficiency since they see the source through the ACS (Fig. 7). Taking into account the presence of the alignment device, the detection plane homogeneity of Fig. 6 is understood: the alignment device shadows detector 0; *a posteriori* it was verified that the alignment angle α was slightly off the theoretical value of 0°, and therefore it slightly shadowed detectors 2 and 3. This can be seen from the absorption feature at low energy for those detectors. During flight, the alignment device will remain present, and thus complicates the response function of the spectrometer, its Monte-Carlo mass model, and in particular the image de-convolution for off-axis sources.

**3.2. Efficiency and homogeneity measurements in the high-energy domain**

For efficiency and homogeneity measurements above 2 MeV the 4 MV van de Graaf accelerator of the Bruyères le Châtel center was used in order to produce a high-intensity proton beam impinging on a water cooled $^{13}$C target (thickness 100 μg/cm$^2$). The $^{13}$C(p,γ)$^{14}$N nuclear reaction in this target produces an excited state of $^{14}$N, which decays and generates a set of photons of well-known energies. This reaction has two main resonances at proton energies of 550 keV and 1742 keV, with quite different characteristics. During a first 36 h data acquisition period, protons of 1742 keV (current 100 μA) were used to generate photons of up to 9169 keV, with narrow lines used for a precise energy calibration. In 2 more periods of 36 h, the proton beam energy was set to 565 keV (current 250 μA) and a different set of photons of up to 8062 keV was generated, with broader lines but higher statistics, used for efficiency measurements. For these measurements, SPI was located at 8 m from the target under an angle of +45° away from the beam. The SPI mask was removed and the target was on the SPI optical axis (for the last 36 h period, SPI was turned by 8°). A glass window separated the accelerator hall from the clean room. A reference Ge detector, used for monitoring the accelerator photon production, was located at 1 m from the target, under an angle of –45° from the beam.

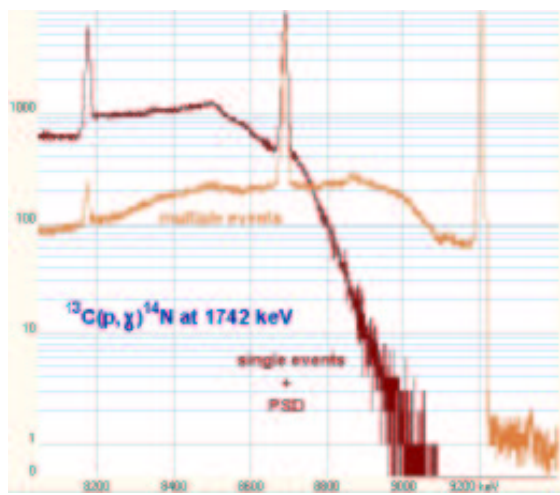

**Figure 8.** Spectra acquired with SPI during the run with 1742 keV protons.

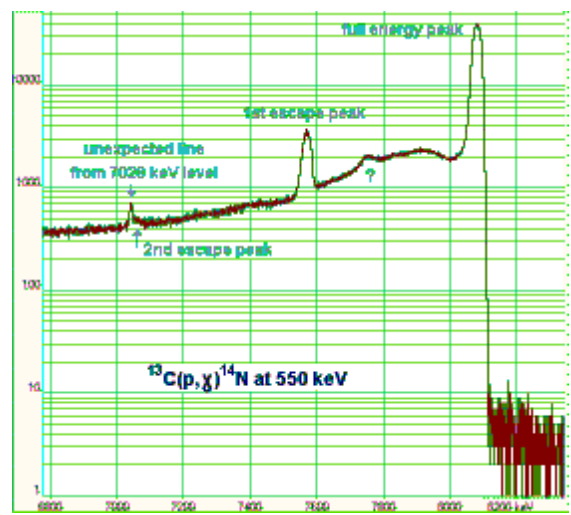

**Figure 9.** Spectra acquired with SPI during the runs with 565 keV protons.

Fig. 8 shows the high energy end of the spectrum acquired with SPI during the 1742 keV proton energy run with the ACS switched off. The 9198 keV line is outside the high energy range of a Ge-detector preamplifier, such that it is not seen by SPI in the single detector event tables (SE+PE), but can be seen for photons which split their energies among multiple detectors (ME table). For some of the 19 Ge detectors, the high-energy range extends up to 8.8 MeV, such that they detect the 1$^{st}$ escape peak (at 8687 keV). All 19 Ge detectors see the 2$^{nd}$ escape peak (at 8176 keV). This narrow line, visible because the ACS has been switched off, is used for the calibration of the energy vs. ADC-bin relation in the high-energy domain. The nuclear reaction at 1742 keV is not used for efficiency studies, since (as discussed below) the production of photons by the reaction at 1742 keV is very non isotropic, and therefore it is difficult to predict the photon flux and the branching ratios for intermediate energy levels.

On the contrary, the 550 keV resonance is very large (23 keV width), it is considered as almost isotropic and gives high fluxes. Fig. 9 shows the high energy end of the spectrum observed by SPI (with ACS switched on), with the 8062 keV line visible and their 1$^{st}$ and 2$^{nd}$ escape peaks. Superimposed on the 2$^{nd}$ escape peak, an unexpected narrow line from the 7028 keV level can be observed which is not mentioned in the decay schemes of the $^{14}$N nucleus for this resonance.

However the energy of the lines obtained on a thick target can not be precisely determined. We used thick targets of $^{13}$C (100 µg/cm$^2$) in order to increase the interaction probability (reaction rate). In such a target a 565 keV proton can loose 34 keV. According to the depth at which the interaction takes place inside the target, the energy of the collision varies and so does the photon energy. Therefore the line profile is some convolution of the resonance profile and energy loss of protons in the target. Fig. 10-12 show the 8062 keV line observed with the Ge-monitor for different proton beam energies and target thicknesses. Furthermore the real proton energy is not well determined because it depends on the beam tuning. Thus this resonance is not used for energy calibration, but for efficiency studies only.

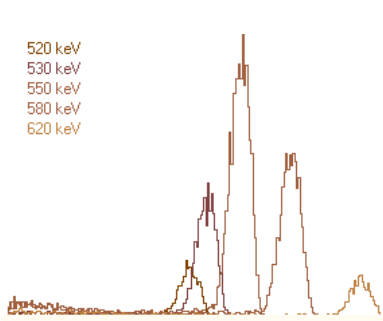

**Figure 10.** Line profiles for a 40 µg/cm$^2$ target for protons from 520 to 620 keV. The 550 keV resonance profile is scanned.

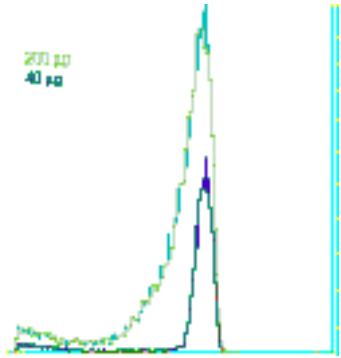

**Figure 11.** Line profiles for 550 keV protons on 40 and 200 µg/cm$^2$ targets.

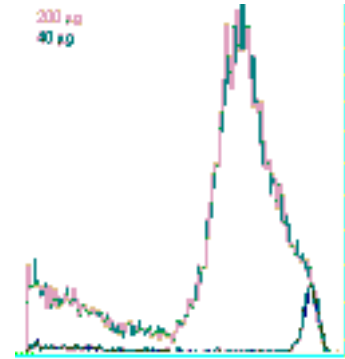

**Figure 12.** Line profiles for 620 keV protons on 40 and 200 µg/cm$^2$ targets.

By integrating the counts in the lines, and knowing the branching ratios towards lower energy levels, the SPI Ge-detector efficiency for the high-energy domain can be extracted. Over 1.7 million photons are available in the 8062 keV line for analysis, which is in progress.

The 1742 keV resonance of the reaction shows interesting phenomena. This resonance excites the 9172 level in the resulting $^{14}$N nucleus. This nucleus decays through a cascade of different levels, with γ-rays from 9169 down to 972 keV. The 9172-9169 keV difference is due to the translation from the center of mass of the (proton,$^{13}$C) system to the laboratory rest frame. Due to the collision, the nucleus moves in the target. Depending on the lifetime of its starting level, the γ-photon is emitted with some Doppler shift, which is a function of the angle between the nucleus velocity and the observer. It is zero for 90° and maximal on the beam axis (forward-emitted photons are blue shifted). It also depends on the slowing-down process of the nucleus in the target, giving rise to distorted line profiles. With a long enough lifetime of the starting level, it's even possible to observe a part of the photons emitted at rest, without Doppler shift. Fig. 13 shows a set of lines with different life times and Doppler profiles, obtained in March 2002 with the Ge-monitor placed at 0°, 45° and 90° with respect to the proton beam. The 6444 keV line originates from the 6446 keV level (lifetime 430 fs). The 6538 keV line (1$^{st}$ escape of the 7028 keV line) originates from the 7029 keV level (lifetime 3.7 fs). In the latter case, the excited nucleus has not enough time to stop before the level decay. All these effects apply also to the 550 keV resonance, but masked by the widths of the lines.

Fig. 13 shows an interesting feature, the photon emission anisotropy for different transitions between levels. The ratio of the line intensities at various angles are quite different: the 6368 keV line (1$^{st}$ escape of the 6857 line) has its maximum at 45° and is practically not visible at 90° (respectively 947, 2519 and 271 photons for 0°, 45° and 90°). On the contrary, the 6538 keV line has its minimum at 0° (respectively 1581, 4672 and 3530 photons). Therefore the branching ratio found in the literature cannot be applied directly to this highly anisotropic resonance.

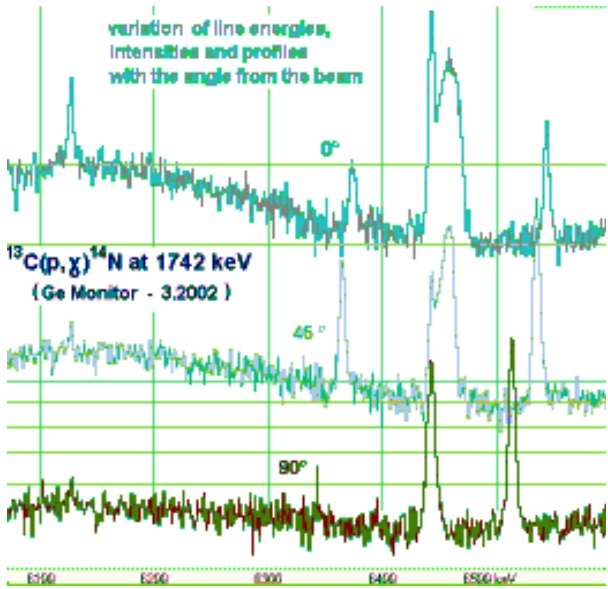

**Figure 13.** Line profiles and intensities as a function of level lifetime and observation angle, measured with a Ge-monitor, March 2002.

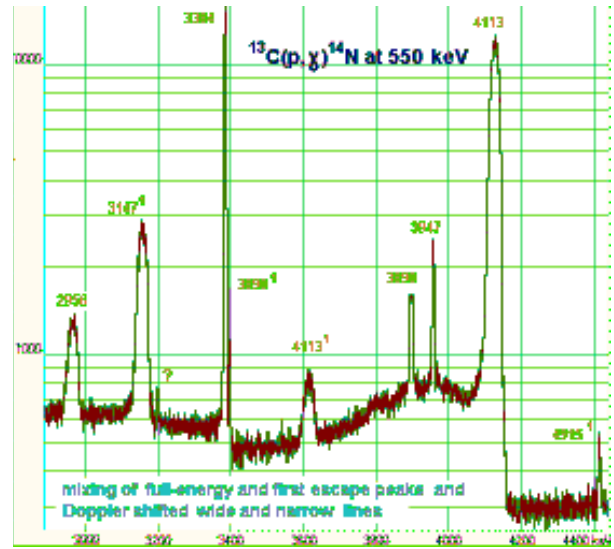

**Figure 14.** Mixing of full-energy and first escape peaks, Doppler shifted wide and narrow lines observed with SPI for the 550 keV resonance. Line identification is not trivial.

**3.3. Imaging performance measurements with mask and sources at 125 m**

The measurements of the imaging performance were performed with the mask mounted, using high-activity radioactive sources placed at 125 m from SPI on an alignment hardware, located outside the experimental clean room, which remained confined by a plastic window. The source distance of 125 m ensures a nearly parallel beam, such that the source appears to SPI as point-like. The photon beam was defined by collimators, designed in such a way that the beam diameter was about 4 m at SPI level. To verify the source alignment, a monitoring tool, made of a NaI detector mounted on a motorized XY-scanning device, checked the horizontal and vertical beam profile before each SPI data acquisition.

For an acceptable counting rate at the detector, a high source activity is needed. With the following 4 different high-activity sources a large energy domain was covered: $^{241}$Am (activity 3 Ci, photopeak energy 60 keV), $^{137}$Cs (0.5 Ci, 661 keV), $^{60}$Co (0.25 Ci, 1172 keV, 1332 keV) and $^{24}$Na (0.08 Ci, 1370 keV, 2753 keV). The $^{24}$Na source with its short lifetime of about 15 hours was specially produced for this purpose in a nuclear reactor of CEA Saclay (France), then quickly transported to the Bruyères le Châtel calibration site. Each source was shielded from the environment using a massive container. For security reasons these sources were operated at night, when less people were present on the site, by a team of specialists from DIMRI (CEA/Saclay).

Fig. 15 shows as an example the pattern of the counts on the germanium camera in the 2.75 MeV photopeak for the high activity $^{24}$Na radioactive source placed at 125 m from SPI. Fig. 16 shows the corresponding reconstructed image in the 2.75 MeV line. It is the highest-energy image so far obtained by a mask-projection technique on a space platform. A detailed imaging analysis of the Bruyères calibration results is presented in Ref. 12 and confirms that the resolving power for two sources obtained with image deconvolution tools is better than 1° at 1 MeV (for high signal-to-noise ratios) and the point-source location is better than 10' (under the same conditions).

Interleaved with the imaging performance measurements, during daytime low-intensity 8 m sources were used to illuminate the spectrometer in its flight configuration with the mask present. In dedicated acquisition periods the optical axis of SPI was oriented with respect to each source at different angles (varying between 0° and 360°). Incidence angles outside the field of view of the telescope are important for ACS efficiency and leakage studies.

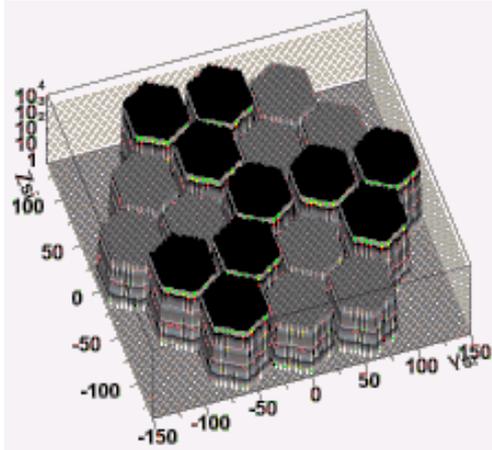 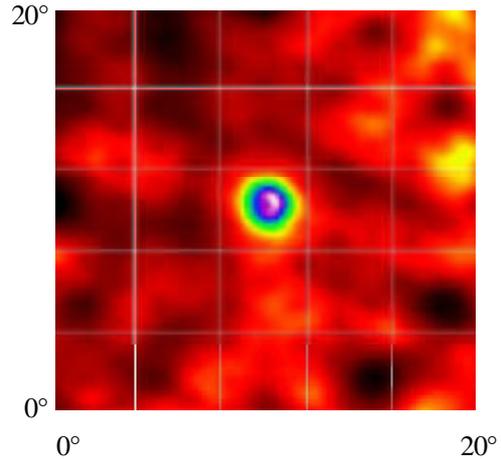

**Figure 15.** The mask pattern is reflected by the pattern of the germanium detector counts produced by a $^{24}$Na source at long distance and aligned on the optical axis of the spectrometer (2.75 MeV peak).

**Figure 16.** Reconstructed image of the 2.75 MeV line for the high activity $^{24}$Na source after 19 pointings (IROS). An angular resolution better than 2° (FWHM) is obtained with the mask imaging technique of SPI.

## 4. CROSS-CALIBRATION OF SPI AND IBIS

A calibration campaign with the completely integrated INTEGRAL satellite was performed at the ESA center of Noordwijk during three weeks in January/February 2002. This calibration phase was the main on-ground calibration for IBIS and JEM-X, which were leading the tests, while for SPI it was complementary to the Bruyères le Châtel campaign. During the measurements INTEGRAL was operated vertically in the ESA clean room, and an irradiation device was installed on a horizontally movable crane at 9.3 m above the detection plane of SPI. The irradiation device, built by CEA/Saclay, could be used in two modes. In the first mode, for a given measurement a radioactive source was placed in a source holder already used for the Bruyères le Châtel measurements. The source holder was then carried by an operator via an elevator 12 m above ground and screwed into its corresponding source collimator on the irradiation device. The irradiation device was then positioned over the satellite in its measurement position (on- or off-axis) over the instruments by motion of the crane. The second mode of operation used an X-ray generator, operated at up to 120 kV, mounted on the irradiation device. It could directly illuminate the satellite with a smooth photon spectrum up to 120 keV, or illuminate a target for production of X-ray fluorescence lines. Targets used were Nd (lines at 37.1 keV and 42.6 keV), Ag (lines at 22.0 keV and 24.9 keV) and Mo (17.3 keV and 19.5 keV). W and U targets were also foreseen but not used. The X-ray generator was calibrated for these targets in Saclay using a monitoring detector.

The use of the X-ray generator extends the energy domain down to 17.3 keV, compared to 60 keV in Bruyères le Châtel. Fig. 17 shows the spectrum obtained with SPI for the Nd target (with a peak FWHM of 2.1 keV for 37.1 keV photons, with cooled detectors but preamplifiers operated at ambient temperature). The high-energy domain was explored up to 2.75 MeV using a $^{24}$Na source produced for this purpose in the CEA/Saclay nuclear reactor and shipped to Noordwijk on the same day. Fig. 18 shows the corresponding photopeak.

During the Noordwijk calibration a total of 114 data-acquisition periods (runs) in different configurations were recorded, they are currently being analyzed. The Noordwijk measurements permit crosschecks with the Bruyères le Châtel calibration results through the reuse of a part of the radioactive sources already used for the 8 m short-distance-source measurements in Bryuères. Image-performance verifications are possible since the spacecraft was in its flight configuration with the mask mounted. But the analysis is more complex than for the Bruyères data, where a near parallel beam has been produced with the high-intensity sources placed at 125 m. Imaging verification would require the computation of the instrument-response function for the Noordwijk source geometry.

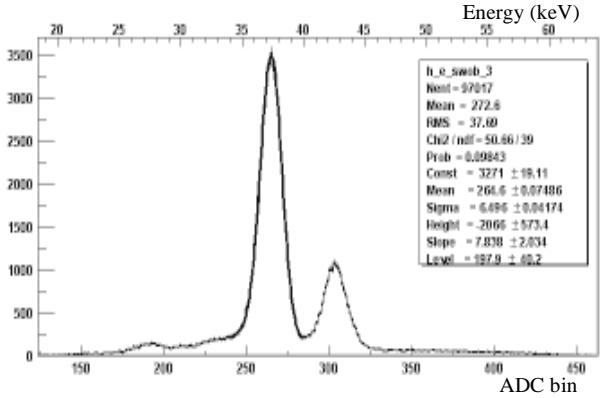 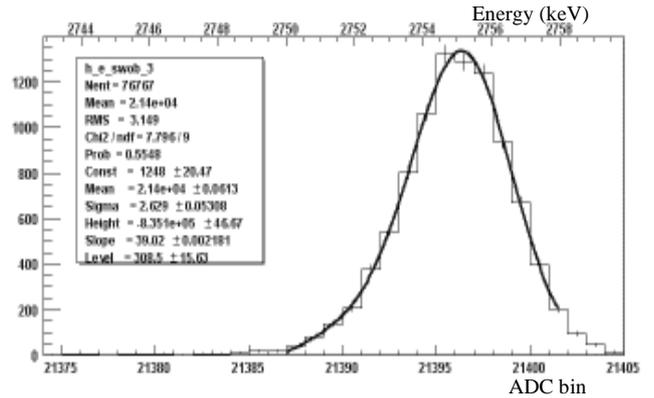

**Figure 17.** SPI detection of the X-ray generator spectrum produced by X-ray fluorescence on a Nd target at Noordwijk. The two 37.1 keV and 42.4 keV lines are well separated by one SPI Ge detector, with a FWHM of 2.1 keV (with hot preamplifiers).

**Figure 18.** SPI detection of the photopeak from the 2.75 MeV line produced by a $^{24}$Na source used at the Noordwijk INTEGRAL instruments cross-calibration, with a FWHM of 3.4 keV (with hot preamplifiers).

For SPI, in addition to the extension of the energy domain to low energies, two important aspects of those measurements are the study of the background with SPI integrated on the satellite, and the study of the influence of other satellite components which act as absorbers to photons on their path to the SPI detector. In particular, special source positions have been used in order to study the projections of the IBIS and JEM-X masks onto the SPI detection plane. While the SPI mask projection, with its coarse opaque/transparent structure, onto the IBIS detection plane can be seen directly by its pattern left on the fine-segmented IBIS detectors (Fig. 19), a detailed INTEGRAL-level Monte-Carlo model is necessary in order to understand the counting rates observed by the projection of the fine-segmented IBIS or JEM-X masks on each of the 19 Ge detectors of the SPI detection plane.

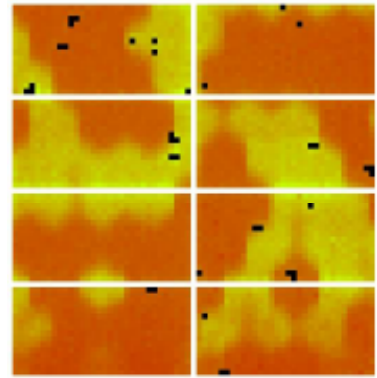

**Figure 19.** Count-image recorded by the 4096 pixels of the high-energy detector of IBIS (PICSIT) for a run where the $^{24}$Na source, the SPI mask and the IBIS detector were aligned. In this configuration a part of the SPI coded mask pattern is projected on the IBIS detection plane. Image deconvolution using the SPI mask becomes possible with IBIS, thus increasing the IBIS coded field-of-view (at the cost of noisier images for on-axis sources before background subtraction). A detailed INTEGRAL-level mass model is necessary in order to account for sources located in the extended part of the IBIS coded field-of-view.

## 5. CONCLUSIONS

The results obtained during the on-ground calibration confirmed the expected detector performance. The understanding of multiple-detector interactions and the impact on the anti-coincidence system have to be refined. Full offline analysis of all measurements is in progress at the SPI collaboration institutes. One of the goals is to verify the response simulation based on the mass-model of the instrument, in particular for understanding off-axis measurements.

For all INTEGRAL instruments extensive on-ground calibration campaigns were performed in order to understand the coded-aperture telescopes before launch. During the first three months after launch in October 2002, an in-flight performance verification phase is foreseen. The initial out-gassing period is followed by the step-by-step instrument switch-on, an instrument-configuration tuning phase, and an in-flight calibration with the observation of astrophysical sources like Cygnus X-1 and the Crab nebula. Those data will be complementary to the on-ground calibration data and the INTEGRAL simulations. The joint understanding of all three data sets will contribute to the success of the mission.

## ACKNOWLEDGMENTS

We would like to thank the staff of the CEA center in Bruyères le Châtel (France), as well as the staff of the ESA/ESTEC center in Noordwijk (The Netherlands) and the Alenia team from Turin (Italy), for their help and hospitality during the successive SPI calibration activities, as well as the staff from CNES in Toulouse (France) for their generous support during installation and operation of the spectrometer during all test phases. Many thanks to the DIMRI staff from CEA/Saclay for the handling of the high activity radioactive sources during the calibration phases and the precious help during preparation of the calibration campaigns. Thanks go also to the IBIS team for Fig. 19.